\documentclass[epj,nopacs]{svjour}
\journalname{Journal}

\usepackage{amsmath}
\usepackage{amssymb}
\usepackage{isomath}
\usepackage{hyperref}
\makeatletter
\def\cl@chapter{\@elt {theorem}}
\makeatother
\expandafter\def\csname ver@luatexbase.sty\endcsname{}
\usepackage{cleveref}
\crefname{section}{sect.}{sects.}
\usepackage[sort&compress,numbers]{natbib}
\usepackage{jabbrv}
\DefineSpuriousJournalWord{The}
\DefineSpuriousJournalWord{for}
\DefineSpuriousJournalWord{on}
\DefineJournalException{Computers \& Fluids}{Comput. Fluids}
\DefineJournalException{Computers \& Mathematics with Applications}{Comput. Math. Appl.}
\DefineJournalException{Journal de Physique II}{J. Phys. II}
\DefineJournalException{Physica D: Nonlinear Phenomena}{Physica D}
\DefineJournalException{Progress in Computational Fluid Dynamics, an International Journal}{Prog. Comput. Fluid Dyn.}
\DefineJournalAbbreviation{Non-Newtonian}{Non-Newton}

\IfFileExists{.git/modules/bibtex/hooks/commit-msg.sample}{
    \usepackage{tikz}
    \usepackage{pgfplots}
    \pgfplotsset{compat=1.13}
    \usepackage{gnuplot-lua-tikz}
    \usetikzlibrary{external}
    \usepgfplotslibrary{external}
    \tikzexternalize[prefix=tikz/]

    \usepackage{todonotes}
    \overfullrule=1mm
}{
    \usepackage{tikzexternal}
    \tikzexternalize
    \IfFileExists{tikz/ldc.pdf}{
        \tikzsetexternalprefix{tikz/}
    }{
        \relax
    }
    \let\tikzexternaldisable\relax
    \let\tikzexternalenable\relax

    \newcommand{\todo}[2][]{}
}

\makeatletter
\def\input@path{{./}{plots/}}
\makeatother
\graphicspath{{./}{plots/}}

\makeatletter
\@ifpackageloaded{tikz}{
    \renewcommand{\todo}[2][]{\tikzexternaldisable\@todo[#1]{#2}\tikzexternalenable}
}{
    \relax
}
\def\NAT@spacechar{~} 
\makeatother

\renewcommand{\vec}[1][]{\vectorsym}
\renewcommand{\tens}[1][]{\tensorsym}
\newcommand{\unitvec}[1]{\hat{\vec{#1}}}
\newcommand{\agrid}{\Delta x}
\newcommand{\tgrid}{\Delta t}
\newcommand{\totalviscosity}{\eta}
\newcommand{\viscosityfraction}{\beta}

\begin{document}

\titlerunning{An extensible lattice Boltzmann method for viscoelastic flows}
\title{\the\titlerunning: complex and moving boundaries in Oldroyd-B fluids}

\author{Michael Kuron\inst{1} \and Cameron Stewart\inst{1} \and Joost de Graaf\inst{2} \and Christian Holm\inst{1}}
\institute{Institute for Computational Physics, University of Stuttgart, Allmandring~3, 70569 Stuttgart, Germany, \email{mkuron@icp.uni-stuttgart.de} \and Institute for Theoretical Physics, Center for Extreme Matter and Emergent Phenomena, Utrecht University,
  Princetonplein~5, 3584~CC Utrecht, The Netherlands}

\date{\today}

\abstract{
Most biological fluids are viscoelastic,
meaning that they have elastic properties in addition to the dissipative properties found in Newtonian fluids.
Computational models can help us understand viscoelastic flow,
but are often limited in how they deal with complex flow geometries and suspended particles.
Here, we present a lattice Boltzmann solver for Oldroyd-B fluids that can handle arbitrarily-shaped fixed and moving boundary conditions,
which makes it ideally suited for the simulation of confined colloidal suspensions.
We validate our method using several standard rheological setups,
and additionally study a single sedimenting colloid,
also finding good agreement with literature.
Our approach can readily be extended to constitutive equations other than Oldroyd-B.
This flexibility and the handling of complex boundaries holds promise for the study of microswimmers in viscoelastic fluids.
}

\maketitle

\section{Introduction}

Recent years have seen a surge of interest in the study of viscoelastic fluids,
due to increased experimental understanding and several intriguing results that were obtained in these media.
In particular, microswimmers in viscoelastic fluids show a richer set of behaviors than possible in simple (Newtonian) fluids, which include:
the self-propulsion of a microswimmer with a single hinge \cite{qiu14a,normand08a},
which is forbidden in a Newtonian fluid at low Reynolds number by \citeauthor{purcell77a}'s scallop theorem \cite{purcell77a};
enhanced rotational diffusion of thermophoretic Janus swimmers,
due to time-delayed translation-rotation coupling in polymer
suspensions \cite{gomezsolano16a};
a peak in the motility of \emph{Escheria coli} bacteria as a function of the polymer concentration and thus complexity of the fluid \cite{martinez14a};
and a fundamental change in the way a microorganism propels in response to the rheology of the medium \cite{li17a}.
With the majority of industrially and biologically relevant fluids being viscoelastic \cite{larson99a,peker08a},
many more such surprises lie ahead of us.

This has motivated the development of a wide range of theoretical and numerical methods.
However, solving the associated hydrodynamic problem remains an open challenge,
both in terms of efficiency and in defining the relevant constitutive equations. 
Much of the numerical work has focused on well-established, albeit basic, models of complex media,
such as polymeric fluids described by Oldroyd-B \cite{oldroyd50a} and FENE-P \cite{peterlin66a,bird80a}.
Examples of such solvers applied to microfluidic problems include
the finite volume method \cite{zhang18a,de17a},
the finite element method \cite{zhu12b,li16a},
multi-particle collision dynamics (MPCD) \cite{toneian19a,sahoo19a},
dissipative particle dynamics \cite{tenbosch99a},
the immersed boundary method \cite{li17a},
smoothed-particle hydrodynamics \cite{vazquez19a,king20a-pre},
as well as explicit-polymer models based on Stokesian dynamics \cite{townsend18a} and MPCD \cite{qi20a}.
The open problem is how to simulate a fluid with a well-defined rheological response,
while also allowing for the incorporation of colloidal particles.
Lattice Boltzmann \cite{mcnamara88a,higuera89a,kruger17a} (LB) methods hold particular promise to achieve this goal
due to their computational efficiency \cite{bauer20a} and facile boundary \cite{zou97a} and particle coupling \cite{duenweg09a,ahlrichs98a,ladd94a,aidun98a}, as has been demonstrated in Newtonian media.
A wide variety of viscoelastic LB schemes have been conceived over the years \cite{giraud97a, giraud98a, ispolatov02a, li04a, frantziskonis11a, frank05a, frank06a, dellar14a, onishi05a, osmanlic16a, karra07a, su13b, malaspinas10a, su13a}.
However, despite this long history, which we will summarize in \cref{subsec:history},
there remain multiple unresolved issues,
especially with regard to boundary conditions.

In this paper, we address the issues of simulation of a viscoelastic fluid using LB with arbitrarily-shaped, moving boundaries.
Our method is inspired by the \citet{su13b} algorithm for an Oldroyd-B fluid,
which we re-derive as a finite volume scheme similar to that of \citet{oliveira98a}.
This ensures momentum conservation
and allows us to introduce a boundary coupling that makes no assumptions on the stress at the boundary.
Compared to the LB schemes described in the literature,
further advantages include low memory usage and the absence of unphysical diffusion terms.
After summarizing the relevant theory in \cref{sec:theory} and laying out our numerical method in \cref{sec:method},
we benchmark our algorithm using several standard rheological tests:
time-dependence of the planar Poiseuille flow in \cref{subsec:poiseuille},
steady shear flow in \cref{subsec:couette},
the instabilities in lid-driven-cavity flow in \cref{subsec:ldc},
and extensional flow in the four-roll mill in \cref{subsec:4rm}.
Next, we examine the effect of the coupling of translation and rotation on the sedimentation of a sphere in \cref{subsec:sphere},
showing that we reproduce the shear-induced speed-up.
\IfFileExists{snowman-squirmer.tex}{
	We do the same for two connected spheres in \cref{subsec:snowman}
	and a squirmer in \cref{subsec:squirmer}.
}{}
We discuss our findings and conclude with an outlook on future applications in \cref{sec:summary}.

\section{Theory}
\label{sec:theory}

In this section, we summarize the equations underlying viscoelastic flow problems.
They are commonly split into a Newtonian part and an additional constitutive equation, which describes the stress evolution.
In terms of notation, bold symbols denote vectors $(\vec{Z})_i=Z_i$ and bold sans-serif symbols denote tensors $(\tens{Z})_{ij}=Z_{ij}$.

\subsection{Generalized Stokes equation}
\label{subsec:stokes}

The micro-scale flows under consideration take place at low Reynolds numbers,
so the hydrodynamics are governed by the time-independent Stokes equations,
\begin{align}
\sum\limits_{k=1}^d\frac{\partial}{\partial r_k}\sigma_{ki}(\vec{r},t)&=-F^\text{ext}_i(\vec{r},t),
\label{eq:stokes-general} \displaybreak[0]\\
\sum\limits_{k=1}^d\frac{\partial}{\partial r_k}u_k(\vec{r},t)&=0.
\label{eq:incompressibility}
\end{align}
The first equation corresponds to momentum conservation, and the second equation is the incompressibility condition.
$d$ is the number of spatial dimensions,
and $\vec{u}$ and $\tens{\sigma}$ are the fluid's flow velocity and stress at position $\vec{r}$ and time $t$.
$\vec{F}^\text{ext}$ is a force applied to the fluid.

A Newtonian fluid's stress $\tens{\sigma}$ consists of a viscous stress $\tens{\varepsilon}$ and a pressure $p$:
\begin{align}
	\sigma_{ij}(\vec{r},t)&=\varepsilon_{ij}(\vec{r},t)-p(\vec{r},t)\delta_{ij}, \label{eq:stress} \displaybreak[0]\\
	\varepsilon_{ij}(\vec{r},t)&=\eta_\text{n}\left(\frac{\partial}{\partial r_i}u_j(\vec{r},t)+\frac{\partial}{\partial r_j}u_i(\vec{r},t)\right),
\end{align}
which simplifies \cref{eq:stokes-general} to
\begin{align}
\eta_\text{n}\sum\limits_{k=1}^d\frac{\partial^2}{\partial r_k^2}u_i(\vec{r},t)&=\frac{\partial}{\partial r_i} p(\vec{r})-F^\text{ext}_i(\vec{r},t).
\label{eq:stokes}
\end{align}
$\eta_\text{n}$ is the viscosity of the Newtonian fluid.

The more general case of non-Newtonian fluids adds an extra stress $\tens{\tau}$ to \cref{eq:stress}.
$\tens{\tau}$ evolves according to a constitutive equation.
Its effect on the flow may be absorbed into \cref{eq:stokes}'s force via
\begin{equation}
	\vec{F}^\text{p}(\vec{r},t)=\sum\limits_{j=1}^d \unitvec{e}_j \sum\limits_{i=1}^d \frac{\partial}{\partial r_i} \tau_{ij}(\vec{r},t),
	\label{eq:dtau}
\end{equation}
where $\unitvec{e}_i$ is the $i$-th unit vector.
The total force $\vec{F}^\text{ext}=\vec{F}+\vec{F}^\text{p}$ is a sum of an applied force and the force resulting from viscoelastic stress.

\subsection{Oldroyd-B fluids}
\label{subsec:oldroydb}

There are many different constitutive equations that describe the wide range of complex fluids encountered in applications.
These include Oldroyd-B \cite{oldroyd50a}, Jeffreys \cite{jeffreys76a,bird87b}, Giesekus \cite{giesekus82a}, FENE-P \cite{peterlin66a}, FENE-CR \cite{chilcott88a}, or Phan-Thien-Tanner \cite{phanthien77a}.
For simplicity's sake and because it is widely studied, we will focus on Oldroyd-B.
We will later indicate how our method can be extended to some of the above more realistic models.
Oldroyd-B's $\tens{\tau}$ corresponds to the conformation tensor of the constituent polymers, averaged over a small control volume \cite{deville12a}.
It makes several simplifying assumptions about the fluid,
including that it is made up of dumbbell polymers with zero equilibrium length
and that these are very dilute \cite{bird87a},
to arrive at the following constitutive equation:
\begin{align}
	\frac{\partial}{\partial t} \tau_{ij}(\vec{r},t) =&
	-\sum_{k=1}^d u_k(\vec{r},t)\frac{\partial}{\partial r_k} \tau_{ij}(\vec{r},t) \nonumber\\
	&+ \sum\limits_{k=1}^d \tau_{ik}(\vec{r},t) \frac{\partial}{\partial r_k} u_j(\vec{r},t) \nonumber\\
	&+ \sum\limits_{k=1}^d \tau_{kj}(\vec{r},t) \frac{\partial}{\partial r_k} u_i(\vec{r},t) \nonumber\\
	&+\frac{\eta_\text{p}}{\lambda_\text{p}}\left(\frac{\partial}{\partial r_i}u_j(\vec{r},t) + \frac{\partial}{\partial r_j}u_i(\vec{r},t)\right) \nonumber\\
	&- \frac{1}{\lambda_\text{p}}\tau_{ij}(\vec{r},t)
	.
	\label{eq:oldroydb}
\end{align}
Here, the first term corresponds to advection, the next two terms are due to the polymers being stretched by the velocity gradient, and the final two terms represent the polymer relaxation.
$\lambda_\text{p}$ is the relaxation time of the polymers, while $\eta_\text{p}$ refers to the viscosity added to the fluid by their presence.
For use with the finite volume scheme in \cref{subsec:fvm}, flux and source terms are identified in order to re-cast the equation as a conservation law:
\begin{align}
	\frac{\partial}{\partial t}\tau_{ij}(\vec{r},t) =& -\frac{\partial}{\partial r_k} J_{ijk}(\vec{r},t) + S_{ij}(\vec{r},t) \displaybreak[0] \label{eq:continuity}\\
	J_{ijk}(\vec{r},t)=&\ u_k(\vec{r},t)\tau_{ij}(\vec{r},t) \displaybreak[0] \label{eq:flux}\\
	S_{ij}(\vec{r},t)=&
	\phantom{+}\tau_{ij}(\vec{r},t)\underbrace{\sum_{k=1}^d\frac{\partial}{\partial r_k} u_k(\vec{r},t)}_{=0 \text{ per \cref{eq:incompressibility}}} \nonumber\\
	&+ \sum\limits_{k=1}^d \tau_{ik}(\vec{r},t) \frac{\partial}{\partial r_k} u_j(\vec{r},t) \nonumber\\
	&+ \sum\limits_{k=1}^d \tau_{kj}(\vec{r},t) \frac{\partial}{\partial r_k} u_i(\vec{r},t) \nonumber\\
	&+\frac{\eta_\text{p}}{\lambda_\text{p}}\left(\frac{\partial}{\partial r_i}u_j(\vec{r},t) + \frac{\partial}{\partial r_j}u_i(\vec{r},t)\right) \nonumber\\
	&- \frac{1}{\lambda_\text{p}}\tau_{ij}(\vec{r},t) \label{eq:source}
	.
\end{align}

\subsection{Dimensionless numbers}
\label{subsec:dimensionless}

It is common practice in fluid mechanics to introduce certain dimensionless numbers.
Many phenomena do not depend on precise parameter values, but rather on the relative significance of individual physical effects.
The Reynolds number gives the ratio of inertial forces to viscous forces:
\begin{equation}
	\mathrm{Re}=\frac{\rho U L}{\totalviscosity},
\end{equation}
where $L$ is a characteristic length scale of the flow and $U$ a characteristic velocity.
$\mathrm{Re}$ represents the relative importance of inertia.
The Stokes \cref{eq:stokes-general} is only valid in the limit of $\mathrm{Re} \ll 1$.
The Deborah number is determined by the ratio of the elastic relaxation time to the characteristic time scale of the flow \cite{dealy10a}:
\begin{equation}
	\mathrm{De}=\frac{\lambda_\text{p}U}{L},
\end{equation}
thus representing the degree of elasticity in response to a deformation.
The Weissenberg number relates the elastic relaxation time to the characteristic rate at which the deformation is driven \cite{dealy10a}:
\begin{equation}
	\mathrm{Wi}=\lambda_\text{p}\dot{\gamma}.
\end{equation}
Finally, it is convenient to introduce the polymer viscosity fraction
\begin{equation}
	\viscosityfraction=\frac{\eta_\text{p}}{\totalviscosity},
\end{equation}
which can easily be varied while keeping the total viscosity
\begin{equation}
	\totalviscosity=\eta_\text{n}+\eta_\text{p}
\end{equation}
constant.

\section{Numerical methods}
\label{sec:method}

Just as the equations in \cref{sec:theory} are split into a Newtonian part and a viscoelastic constitutive equation, we employ two separate numerical methods.
The former is solved via lattice Boltzmann (LB), while the latter uses the finite volume (FV) method.

\subsection{Lattice Boltzmann}
\label{subsec:lb}

LB \cite{mcnamara88a,kruger17a} constructs solutions to \cref{eq:stokes} from the Boltzmann transport equation (BTE), which derives from the same conservation laws.
The BTE describes the time evolution of $f(\vec{r}, \vec{v}, t)$, which is the probability distribution function of finding a single fluid molecule with velocity $\vec{v}$ at position $\vec{r}$ and time $t$.
LB discretizes the BTE on a lattice, typically a square/cubic lattice in two/three dimensions, with grid spacing $\agrid$ and discrete time steps $\tgrid$.
Relaxation of $f$ toward its Maxwellian equilibrium is linearized and only a finite set of velocities $\vec{c}_i$ is permitted to allow probability to be exchanged solely between neighboring cells.
The probability distribution is thus replaced by the populations $f_i(\vec{r},t)=f(\vec{r}, \vec{c}_i, t)$, with their equilibrium values $f_i^\text{eq}(\vec{r},t)$.
We use the D3Q19 velocity set in three dimensions and D2Q9 for two-dimensional systems.
In the general D$d$Q$q$ notation, $d$ refers to the dimensionality and $q$ to the number of velocity vectors pointing to neighbor cells ---
here these are the six face and twelve edge neighbors (or four edge and four corner neighbors in two dimensions).
The employed two relaxation time (TRT) collision operator relaxes symmetric ($+$) and antisymmetric ($-$) linear combinations of $f_i$ separately,
and only the symmetric relaxation time $\lambda_+$ affects the viscosity of the fluid.
$\lambda_-$ can be tuned to improve the accuracy of boundary conditions \cite{ginzburg08a}.

The full LB method is given by
\begin{align}
f_i(\vec{r}+\vec{c}_i\tgrid,t+\tgrid)
&=f_i(\vec{r},t) \nonumber \\
&\phantom{=}
-\lambda_+\left(f_i^+(\vec{r},t)-f_i^{\text{eq}+}(\vec{r},t)\right) \nonumber \\
&\phantom{=}
-\lambda_-\left(f_i^-(\vec{r},t)-f_i^{\text{eq}-}(\vec{r},t)\right) \nonumber \\
&\phantom{=}
+\Delta_i (\vec{r},t)\label{eq:lb} \\
\intertext{with}
f_i^{\pm}(\vec{r},t) &= \frac{1}{2} \left( f_i(\vec{r},t) \pm f_{-i}(\vec{r},t) \right), \displaybreak[0]\\
f_i^{\text{eq}\pm}(\vec{r},t) &= \frac{1}{2} \left( f_i^\text{eq}(\vec{r},t) \pm f_{-i}^\text{eq}(\vec{r},t) \right), \displaybreak[0]\\
f_i^\text{eq}(\vec{r},t) &=  w_i\rho(\vec{r},t) \left( 1 + 3\vec{c}_i\cdot\vec{u}(\vec{r},t) \phantom{\frac{1}{6}} \right. \label{eq:lb-equilibrium} \\
&\phantom{=} \left. + \frac{1}{6}\left(\vec{c}_i\cdot\vec{u}(\vec{r},t)\right)^2 -\frac{1}{6}u(\vec{r},t)^2 \right), \nonumber \displaybreak[0]\\
\eta_\text{n}&= \rho(\vec{r},t)\left(\frac{1}{3\lambda_+}-\frac{1}{6}\right), \displaybreak[0]\\
\lambda_-&=\frac{3}{16\lambda_+},
\end{align}
and $-i$ defined via $\vec{c}_{-i}=-\vec{c}_i$.
The local fluid density $\rho(\vec{r},t)$ appears explicitly because LB does not simulate a perfectly incompressible fluid.
For consistency, we did verify in our simulations that the fluid does not compress appreciably.
The populations $f_i$ and the macroscopic flow fields are connected via
\begin{align}
	\rho(\vec{r},t)&=\sum\limits_{i=1}^q f_i(\vec{r},t), \displaybreak[0]\\
	\vec{u}(\vec{r},t)&=\frac{1}{\rho(\vec{r},t)}\sum\limits_{i=1}^q f_i(\vec{r},t)\vec{c}_i +\frac{1}{2}\vec{F}^\text{ext}(\vec{r},t) \tgrid \label{eq:macroscopic-velocity} .
\end{align}

$\Delta_i(\vec{r},t)$ in \cref{eq:lb} represents the force $\vec{F}^\text{ext}$ applied to the fluid.
One possible expression for it is given by \citeauthor{guo02a}~\cite{guo02a,schiller08a,schiller14a}:
\begin{align}
\Delta_i(\vec{r},t) &= \frac{3w_i\agrid^2\tgrid^2}{\rho} \left[\vec{F}^\text{ext}(\vec{r},t)\cdot\vec{c}_i \phantom{\frac{\tgrid^2}{\agrid^2}}\right. \nonumber \\
& \phantom{++}+ \left.\frac{3}{2} \mathrm{Tr}\left(
\tens{G} \left( \vec{c}_i\otimes\vec{c}_i \right)\frac{\tgrid^2}{\agrid^2} - \frac{1}{3}\tens{G}
\right)
\right] \label{eq:guo} \\
\intertext{with}
\tens{G} &= \frac{2-\lambda_\text{e}}{2} \left( \vec{u}(\vec{r},t)\otimes\vec{F}^\text{ext}(\vec{r},t) \right. \nonumber \\
&\phantom{=\frac{2-\lambda_\text{e}(}{2}}
\left. + \vec{F}^\text{ext}(\vec{r},t)\otimes\vec{u}(\vec{r},t) \right),
\end{align}
where $w_i$ is the lattice weight factor for $\vec{c}_i$,
$\cdot$ is the scalar\slash dot product,
$\otimes$ is the tensor\slash dyadic product,
and $\mathrm{Tr}$ is the trace of a tensor.

Velocity boundary conditions can be imposed on the fluid by using
\begin{equation}
f_i(\vec{r}_\text{b}+\vec{c}_i\tgrid,t+\tgrid)
\equiv f_{-i}(\vec{r}_\text{b},t)
+ \frac{6\rho w_i\tgrid^2}{\agrid^2} \vec{c}_i\cdot \vec{u}_\text{b},
\label{eq:ubb}
\end{equation}
where $\vec{r}_\text{b}$ is a boundary node with velocity $\vec{u}_\text{b}$ and $\vec{r}_\text{b}+\vec{c}_i\tgrid$ is a fluid node.
For no-slip conditions $\vec{u}_\text{b}=0$, this scheme corresponds to a bounce-back of the population.

\subsection{Background on viscoelastic LB}
\label{subsec:history}

As early as \citeyear{giraud97a},
\citet{giraud97a,giraud98a} used LB to compute the response of the Jeffreys fluid.
This was followed up by \citet{ispolatov02a},
who employed LB to solve a linear Maxwell model,
by implementing the elastic-stress contribution as a body force onto their fluid. 
Similar approaches were followed by \citet{li04a} and \citet{frantziskonis11a}.
Later, \citet{frank05a,frank06a} went beyond the body-force coupling and introduced the effect of elastic stress directly into the second moment of the equilibrium distribution,
which has recently been revisited by \citet{dellar14a}.
Other coupling forms were considered by \citet{onishi05a} and \citet{osmanlic16a},
who employ a Fokker-Planck-like evolution of microscopic dumbbells in a viscous fluid.
This type of system is theoretically known to result in a viscoelastic response that resembles Oldroyd-B \cite{bird95a}.
More direct approaches to reproducing Oldroyd-B were followed by \citet{karra07a} and \citet{su13b},
who solved the stress evolution equation for the corresponding constitutive relation directly using the LB fluid velocity as
input to a finite difference scheme.
\Citet{malaspinas10a} and \citet{su13a} similarly used an LB scheme as a generic differential equation solver and treated the viscoelastic stress tensor component-wise,
for both the Oldroyd-B and FENE-P constitutive relations.
\Citet{phillips11a} provide a more in-depth review of the cited methods for viscoelastic fluids, as well as methods for generalized Newtonian fluids;
for a discussion of LB methods that deal with viscoelastic behavior of active fluids, see the review by \citet{carenza19a}.

The LB schemes listed above are not applied to problems with boundaries \cite{giraud97a, frank05a},
do not require explicit treatment of the stress \cite{ispolatov02a, li04a, frantziskonis11a, frank06a},
or use bounce-back rules to impose specific boundary conditions on the stress \cite{dellar14a, giraud98a}.
Some extrapolate stress onto boundaries to allow for cases where no analytic expression exists \cite{malaspinas10a,su13a},
while others can only be applied to systems for which the stress at the boundary is known beforehand \cite{su13b}.
In the following, we build upon this body of knowledge
and introduce a general method capable of handling complex and moving boundaries.
By doing so, we overcome the limitations of previous viscoelastic LB algorithms.

\subsection{Finite volume method}
\label{subsec:fvm}

Our method is inspired by an LB-coupled finite volume (FV) solver for the electrokinetic equations \cite{capuani04a}
and has similarities to other FV Oldroyd-B solvers \cite{oliveira98a}.
FV methods \cite{versteeg07a} are well suited for solving problems governed by conservation laws such as \cref{eq:continuity} since they guarantee the conservation of, e.g., momentum and energy to machine precision.
We found that the hybrid scheme of finite differences (FD) and LB originally suggested by \citet{su13b} led to violation of energy conservation in the constitutive equation in the presence of boundaries,
which translated into a violation of the conservation of momentum.
Moving boundary simulations as presented in \cref{subsec:sphere} were therefore impossible.

\subsubsection{Discretization}
\label{subsubsec:discretization}

\Cref{eq:continuity} is averaged over one cell's volume $V=\agrid^d$ with surface unit normal $\unitvec{n}$ to become
\begin{align}
	\frac{\partial}{\partial t} \bar{\tau}_{ij}(\vec{r},t)
	&=
	-\frac{1}{V}\int_V\sum\limits_{k=1}^{d}\frac{\partial}{\partial r_k}J_{ijk}(\vec{r},t)\mathrm{d}V+\bar{S}_{ij}(\vec{r},t) \nonumber \\
	&=
	-\frac{1}{V}\int_{\partial V}\sum\limits_{k=1}^{d}J_{ijk}(\vec{r},t)n_k\mathrm{d}S+\bar{S}_{ij}(\vec{r},t),
\end{align}
where Gau{\ss}'s divergence theorem has been applied and the overbar indicates the volume average.
By locating $\bar{\tens{\tau}}$ and $\bar{\tens{S}}$ at the cell center and $\tens{J}$ between two cells, the discrete form of this equation is obtained as
\begin{align}
	\bar{\tau}_{ij}(\vec{r},t+\tgrid)
	&\approx
	-\frac{1}{V}\sum\limits_{\ell=1}^q \sum\limits_{k=1}^{d} J_{ijk}(\vec{r}+\frac{1}{2}\vec{c}_\ell\tgrid,t)c_{\ell k} \nonumber\\
	&\phantom{=}\,+\bar{S}_{ij}(\vec{r},t) + \bar{\tau}_{ij}(\vec{r},t)
	,
\end{align}
where we have used the same grid spacing and time step as in \cref{subsec:lb}.
The neighbor set $\{\vec{c}_i\}$ does not necessarily need to match the one used in \cref{subsec:lb}:
we have found D3Q27\slash D2Q9 to deliver no appreciable advantage over D3Q7\slash D2Q5 \cite{mazumder15a} and have thus selected the latter for its lower computational cost.

We numerically interpolate $Z\in\{\vec{u},\tens{\tau}\}$ as
\begin{align}
	Z(\vec{r}+\frac{1}{2}\vec{c}_i\tgrid,t) &\approx \frac{1}{2}\left(Z(\vec{r},t) + Z(\vec{r}+\vec{c}_i\tgrid,t)\right)
	\label{eq:interpolation}
\end{align}
and insert these expressions into \cref{eq:flux} to obtain
\begin{align}
	\tens{J}_i&(\vec{r}+\frac{1}{2}\vec{c}_i\tgrid,t)=\nonumber \\
	&\frac{1}{\left|c_i\right|A_0}
	\tens{\tau}(\vec{r}+\frac{1}{2}\vec{c}_i\tgrid,t) \left(\vec{u}(\vec{r}+\frac{1}{2}\vec{c}_i\tgrid,t)\cdot\vec{c}_i\right),
	\label{eq:flux-discrete}
\end{align}
where the projection onto $\vec{c}_i$ and the prefactor
\begin{equation}
	A_0=\frac{1}{2d}\sum\limits_{\ell=1}^q \left|c_\ell\right|.
\end{equation}
account for the case of $q>2d+1$ \cite{capuani04a}.
We will replace \cref{eq:flux-discrete} in \cref{subsubsec:stability} with a different expression to improve numerical stability.

We need to numerically differentiate $\vec{u}$
and average over the volume of one cell:
\begin{align}
	\int_V\frac{\partial}{\partial r_i}\vec{u}(\vec{r},t) \mathrm{d}V &=
	\int\limits_{-\agrid/2}^{\agrid/2} \mkern-10mu\cdots \mkern-10mu\int\limits_{-\agrid/2}^{\agrid/2}
	\left(\vec{u}(\vec{r}+\frac{\agrid}{2}\unitvec{e}_i,t)\right.\nonumber\\
	&\phantom{=}\left.-\vec{u}(\vec{r}-\frac{\agrid}{2}\unitvec{e}_i,t)\right) 
	\frac{\mathrm{d}r_1\cdots\mathrm{d}r_d}{\mathrm{d}r_i} .
\end{align}
Making the central-point approximation
\begin{align}
	\int\limits_{-\agrid/2}^{\agrid/2} \mkern-10mu\cdots \mkern-10mu\int\limits_{-\agrid/2}^{\agrid/2} &
	\vec{u}(\vec{r}\pm\frac{\agrid}{2}\unitvec{e}_i,t)
	\frac{\mathrm{d}r_1\cdots\mathrm{d}r_d}{\mathrm{d}r_i} \nonumber\\
	&\approx \agrid^{d-1} \vec{u}(\vec{r}\pm\frac{\agrid}{2}\unitvec{e}_i,t)
	\label{eq:centralapprox}
\end{align}
and inserting \cref{eq:interpolation} yields the first-order FV discretization
\begin{equation}
	\frac{\partial}{\partial r_i}\vec{u}(\vec{r},t)\approx \frac{1}{2\agrid}\left(\vec{u}(\vec{r}+\unitvec{e}_i,t)-\vec{u}(\vec{r}-\unitvec{e}_i,t)\right),
\end{equation}
which is identical to the corresponding FD scheme.
Inserting it into \cref{eq:source} then yields $\bar{\tens{S}}(\vec{r},t)$.


The force \cref{eq:dtau} is similarly discretized by averaging over the volume of a cell:
\begin{align}
	F^\text{p}_j(\vec{r},t)
	&=
	\frac{1}{V}\int_V\sum\limits_{i=1}^d \frac{\partial}{\partial r_i} \tau_{ij}(\vec{r},t) \mathrm{d}V \nonumber\\
	&=
	\frac{1}{V}\int_{\partial V}\sum\limits_{i=1}^d \tau_{ij}(\vec{r},t) n_i \mathrm{d}S \\
	&\approx \frac{1}{A_0}\sum\limits_{\ell=1}^q\frac{1}{\left|c_\ell\right|}\sum\limits_{i=1}^d \tau_{ij}(\vec{r}+\frac{1}{2}\vec{c}_\ell\tgrid,t) c_{\ell i},
	\label{eq:force}
\end{align}
where $\tau_{ij}(\vec{r}+\frac{1}{2}\vec{c}_i\tgrid)$ can be obtained via \cref{eq:interpolation}.

Boundaries across which no stress is transported can be imposed on the FV scheme by using
\begin{equation}
	\tens{J}(\vec{r}_\text{b}+\frac{1}{2}\vec{c}_i\tgrid)\equiv 0,
\end{equation}
where $\vec{r}_\text{b}$ is a boundary node and $\vec{r}_\text{b}+\vec{c}_i\tgrid$ is a fluid node.
$\tens{\tau}(\vec{r}_\text{b})$ needs to be extrapolated so that the force can continue to be obtained via \cref{eq:force}.
We found constant extrapolation
\begin{equation}
	\tens{\tau}(\vec{r}_\text{b})\equiv \tens{\tau}(\vec{r}_\text{b}+\vec{c}_i\tgrid)
	\label{eq:extrapolation}
\end{equation}
to be sufficient, but linear or quadratic extrapolation could be employed as needed.

\subsubsection{Stability improvements}
\label{subsubsec:stability}

FV and FD schemes are known to exhibit numerical instabilities in certain situations, which result in spatial oscillations or ``wiggles'' \cite{mazumder15a}.
This is a particularly prominent problem in the context of Oldroyd-B as the model's P\'eclet number \cite{versteeg07a}, which relates advective transport to diffusive transport, is infinite due to the absence of a diffusive term in \cref{eq:oldroydb}.
We observed stress wiggles when performing the simulations of \cref{subsec:sphere,subsec:4rm} as described in \cref{subsubsec:discretization}.
Solutions proposed for Oldroyd-B include:
using higher-order differentiation schemes \cite{su13b},
inserting an artificial diffusion term \cite{malaspinas10a},
or storing $\vec{u}$ and $\tens{\tau}$ on two separate grids shifted relative to each other by half a cell \cite{oliveira98a,alves01a}.
These methods increase computational cost, modify the physics of the system, and make the implementation cumbersome, respectively,
so we consider alternative techniques suggested in general FV literature.
These include higher-order interpolation \cite{rhie83a,versteeg07a}
and differentiation \cite{lilek95a} schemes,
as well as upwind schemes \cite{versteeg07a,capuani04a}.

We resorted to the latter and chose an upwind variant called \emph{corner-transport upwind scheme} suggested by refs.~\citenum{capuani04a,colella90a} and employed in our previous work \cite{rempfer16a,kuron16a}.
Upwind schemes calculate advective fluxes like \cref{eq:flux-discrete} not by interpolating quantities to the midpoint between two cells, but by using the quantity from either cell, depending on which way the flow points \cite{versteeg07a}.
Reference~\citenum{capuani04a}'s method is geometrically motivated by virtually displacing a cell at $\vec{r}$ by its velocity $\vec{u}(\vec{r},t)\tgrid$ and calculating the virtual cell's overlap volume with all neighboring cells.
This overlap corresponds to the fraction of $\tens{\tau}(\vec{r},t)$ to be transferred to the respective neighboring cell.
While this in principle results in fluxes in all D3Q27\slash D2Q9 directions, fluxes beyond the D3Q7\slash D2Q5 neighbor set are $\mathcal{O}(u^2)$, making them negligible here.

\subsection{Moving boundaries}
\label{subsec:mb}

One way of coupling particles to an LB fluid is by the moving boundary method.
It was introduced by \citet{ladd94a} and later enhanced by \citet{aidun98a}.
This method is applicable for particles much larger than the size of a grid cell
and considers the cells inside the particle as no-slip conditions in the particle-co-moving frame.
This corresponds to a velocity boundary condition of
\begin{equation}
	\vec{u}_\text{b}(\vec{r}_\text{b},t)=\vec{v}(t)+\vec{\omega}(t)\times(\vec{r}_\text{b}-\vec{r}(t)),
	\label{eq:surface-velocity}
\end{equation}
which can be applied via \cref{eq:ubb}.
$\vec{r}$, $\vec{v}$, and $\vec{\omega}$ are the position, linear, and angular velocity of the particle.
Applying the boundary condition to the fluid transfers linear and angular momentum to the particle, corresponding to a force and torque
\begin{align}
	\vec{F}(t)&=V \sum\limits_{i=1}^{q} \vec{c}_i \left(f_i(\vec{r}_\text{b},t)+f_{-i}(\vec{r}_\text{b}+\vec{c}_i\tgrid,t)\right), \displaybreak[0]\\
	\vec{T}(t)&=V \sum\limits_{i=1}^{q} \left(\vec{r}_\text{b}-\vec{r}\right) \times \vec{c}_i \left(f_i(\vec{r}_\text{b},t)+f_{-i}(\vec{r}_\text{b}+\vec{c}_i\tgrid,t)\right).
\end{align}
The particle trajectory is obtained by summing these forces and torques, along with any externally applied ones, and integrating numerically with a symplectic Euler integrator.

As a particle moves across the lattice, the set of cells it overlaps changes.
When a cell at $\vec{r}_\text{f}$ is converted from fluid to solid, its fluid populations are deleted.
In the reverse case, new fluid populations are created at their equilibrium value, $f_i(\vec{r}_\text{f},t)=f_i^\text{eq}(\vec{r}_\text{f},t)$ from \cref{eq:lb-equilibrium},
whose velocity $\vec{u}_\text{b}(\vec{r}_\text{f},t)$ is given by \cref{eq:surface-velocity}.
Momentum conservation during creation and destruction of populations is ensured by applying a force to the particle
that balances any momentum destroyed or created:
\begin{equation}
	\vec{F}(t)=\pm\frac{V}{\tgrid}\rho(\vec{r}_\text{f},t)\vec{u}(\vec{r}_\text{f},t).
\end{equation}

The moving boundary method has previously been extended to FV schemes \cite{kuron16a,rivas18a},
but only in the context of ion concentrations propagating according to the electrokinetic equations.
In this paper, we take a similar path to apply it to the $\tens{\tau}$ of a viscoelastic medium.
Refs.~\citenum{kuron16a,rivas18a} take precautions to ensure that charge is conserved.
We do the same here to ensure that stress --- whose diagonal elements correspond to stored energy --- is not created or destroyed while cells are converted between fluid and solid.
Refs.~\citenum{kuron16a,rivas18a} further calculate the fraction of a cell that is overlapped by the particle and use that information to smooth out the conversion process,
which they reported to significantly decrease oscillations in the particle's speed.
For the simulations in \cref{subsec:sphere}, we found such smoothing to be unnecessary.

A fluid cell at $\vec{r}_\text{f}$ that is destroyed in front of the particle has its stress distributed among the surrounding $N_\text{f}$ fluid cells as
\begin{equation}
	\tens{\tau}(\vec{r}_\text{f}+\vec{c}_i\tgrid, t+\tgrid) = \tens{\tau}(\vec{r}_\text{f}+\vec{c}_i\tgrid, t) + \frac{1}{N_\text{f}} \tens{\tau}(\vec{r}_\text{f}, t).
\end{equation}
A cell behind the particle that is created with new fluid receives
\begin{equation}
	\tens{\tau}(\vec{r}_\text{f}, t+\tgrid) = \frac{1}{N_\text{f}+1} \sum\limits_{i=1}^q \tens{\tau}(\vec{r}_\text{f}+\vec{c}_i\tgrid, t),
\end{equation}
and the corresponding amount is removed from the neighboring cells:
\begin{equation}
	\tens{\tau}(\vec{r}_\text{f}+\vec{c}_i\tgrid, t+\tgrid)
	= \tens{\tau}(\vec{r}_\text{f}+\vec{c}_i\tgrid, t)
	- \frac{\tens{\tau}(\vec{r}_\text{f}, t+\tgrid)}{N_\text{f}} .
\end{equation}

\subsection{Implementation and extensibility}

The methods described above are implemented using the \textsf{waLBerla} C++ framework \cite{godenschwager13a,bauer20a}.
It allows for efficient and highly parallelized implementation of local algorithms on regular grids
and provides several LB implementations and a rigid-body dynamics module.
The Python module \textsf{pystencils} \cite{bauer19a} can be used to automatically generate code for grid-based algorithms, either for use in Python or for \textsf{waLBerla}.
We have extended it with a generator for finite volume discretizations that automatically derives the expressions in \cref{subsec:fvm} when provided with the Oldroyd-B~\cref{eq:continuity,eq:flux,eq:source}.
By instead supplying, for example, the FENE-P constitutive equation \cite{peterlin66a}, we could simulate that model without writing any additional code.

There are several other fluid dynamics software packages that allow the user to provide such equations
and automatically derive discretizations for them,
e.g. \textsf{Dedalus}~\cite{burns20a} or \textsf{OpenFOAM}~\cite{jasak07a}.
The combination of \textsf{pystencils} and \textsf{waLBerla}, however, is unique in that it allows for arbitrarily-shaped boundary conditions that change over time, which can be put to use for the moving boundaries of \cref{subsec:mb}.
We forgo \textsf{waLBerla} for the two-dimensional simulations,
since they do not require rigid-body dynamics or parallelization,
and run these simulations completely from Python.
In this case, LB is provided by the \textsf{lbmpy} module \cite{bauer20b-pre}.

\section{Validation and results}
\label{sec:validation}

In this section, we solve multiple rheological benchmark systems to verify the correctness of our algorithm and implementation by comparing against results from literature.
We then simulate \IfFileExists{snowman-squirmer.tex}{systems}{a system} involving moving boundary conditions and translation-rotation coupling in order to demonstrate the strength of the method.

\subsection{Time-dependent Poiseuille flow}
\label{subsec:poiseuille}

\begin{figure}[tb]
\centering
\tikzsetnextfilename{poiseuille}
\begin{tikzpicture}
	\fill[pattern=north west lines] (0,-0.5) rectangle (4,0);
	\draw[thick] (0,0) -- (4,0);
	\draw[thick] (0,2) -- (4,2);
	\fill[pattern=north west lines] (0,2.5) rectangle (4,2);
	\draw[|-|] (0.5,0.1) -- node[anchor=north east] {$L$} (0.5,1.9);
	\draw[->] (1.5,1.5) -- node[anchor=south] {$F$} (2.5,1.5);

	\draw[->] (3,0.2) -- (3,0.7) node[anchor=east] {$y$};
	\draw[->] (3,0.2) -- (3.5,0.2) node[anchor=west] {$x$};
	
	\draw[dashed,black!50] (0,1) -- (4,1);

	\begin{scope}[xshift=5cm,yshift=1cm]
		\draw[thick,variable=\y, domain=-1:1] plot ({-\y*\y*2+2},{\y});
		\draw[->] (0,-1) node[anchor=north] {$0$} -- (2,-1) -- (2.5,-1) node[anchor=west] {$u_x$};
		\draw[->] (0,-1) node[anchor=east] {$0$} -- (0,1) node[anchor=south] {$y$} node[anchor=east] {$L$};
		\draw[dashed,black!50] (0,0) -- (2.5,0);
	\end{scope}
\end{tikzpicture}
\caption{
Geometry of the planar Poiseuille flow system.
A force $F$ is applied to a fluid in a periodic channel of width $L$,
which leads to a parabolic profile across the channel for the flow velocity $u_x$ along the channel.
The dashed line indicates where the flow velocity is measured for further analysis.
}
\label{fig:poiseuille}
\end{figure}
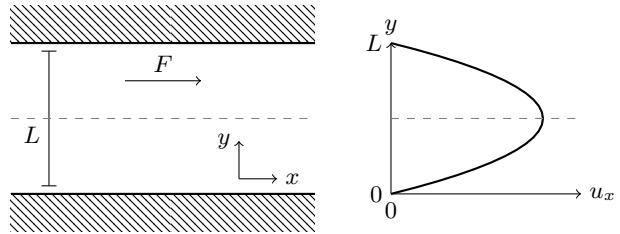

The planar Poiseuille geometry consists of an infinitely long channel of width $L$, through which flow is driven by a homogeneous force along the channel, $\vec{F}=F_x\unitvec{e}_x$.
The channel walls impose a no-slip condition $\vec{u}((x,0)^\intercal,t)=\vec{u}((x,L)^\intercal,t)=0$, while the infinite length can be achieved via periodic boundary conditions in $y$-direction.
This set\-up is illustrated in \cref{fig:poiseuille} and results in a parabolic steady-state flow profile.
Starting this flow in a resting Newtonian fluid causes the steady-state flow to be approached in a monotonous fashion.
In a viscoelastic medium, however, the flow velocity can overshoot its steady-state value and then decay to it on a time scale of $\lambda_\text{p}$.
This is due to the storage of energy in the elastic medium,
which is released back into the fluid on the time scale $\lambda_\text{p}$.
Reference~\citenum{waters70a} provides an analytical expression for the time-dependent velocity at the center of the channel, $\vec{u}((x,\allowbreak L/2)^\intercal,\allowbreak t)$, in a \emph{liquid B'} model.
This model has been shown to be equivalent to Oldroyd-B \cite{xue04a,park09b}.

\begin{figure}[tb]
\centering
\tikzsetnextfilename{poiseuille-eta}
\input{plots/poiseuille-eta.tikz}
\caption{
Velocity at the center of a planar Poiseuille channel over time
for varying viscosity ratios $\viscosityfraction$ and polymer relaxation time $\lambda_\text{p}=3000\tgrid$.
Symbols are numerical calculations, and lines show the analytical prediction by ref.~\citenum{waters70a}.
The solid lines use $L$ from the input parameters, whereas the dashed lines allow it to be a free fit parameter.
}
\label{fig:poiseuille-eta}
\end{figure}

\begin{figure}[tb]
\centering
\tikzsetnextfilename{poiseuille-lambda}
\input{plots/poiseuille-lambda.tikz}
\caption{
Velocity at the center of a planar Poiseuille channel over time
for viscosity ratio $\viscosityfraction=0.3$ and varying polymer relaxation times $\lambda_\text{p}$.
Symbols are numerical calculations, and lines show the analytical prediction by ref.~\citenum{waters70a}.
The solid lines use $L$ from the input parameters, whereas the dashed lines allow it to be a free fit parameter.
}
\label{fig:poiseuille-lambda}
\end{figure}

We choose the channel width $L=28\agrid$, the applied force $F_x=10^{-5}\rho\agrid^4/\tgrid^2$, Newtonian viscosity $\eta_\text{n}=\rho\agrid^2/\tgrid-\eta_\text{p}$, polymer viscosity ratios $\viscosityfraction\in\{0.1,\allowbreak 0.3,\allowbreak 0.5,\allowbreak 0.7,\allowbreak 0.9\}$, and polymer relaxation times $\lambda_\text{p}/\tgrid\in\{1000,\allowbreak 3000,\allowbreak 5000,\allowbreak 7000,\allowbreak 9000\}$ for our test simulations.
This corresponds to a Reynolds number of
\begin{equation}
	\mathrm{Re}
	=\frac{\rho u_x((x,L/2)^\intercal,\infty)L}{\totalviscosity}
	=\frac{\rho F_xL^3}{8\totalviscosity^2}
	= 0.03,
\end{equation}
which is well within the low-Reynolds regime we are interested in.

\Cref{fig:poiseuille-eta} shows the flow velocity $u_x((x,L/2)^\intercal,t)$ over time for various polymer viscosity ratios $\viscosityfraction$ at constant polymer relaxation time $\lambda_\text{p}=3000\tgrid$.
One can see that the magnitude of the overshoot increases with $\viscosityfraction$.
For the largest values of $\viscosityfraction$, the flow can even decay to its final speed in an oscillatory fashion.
\Cref{fig:poiseuille-lambda} keeps $\viscosityfraction=0.3$ constant and varies $\lambda_\text{p}$.
Here it is clear that the magnitude of the overshoot increases with $\lambda_\text{p}$, which is also the characteristic decay time of the overshoot.

\Cref{fig:poiseuille-eta,fig:poiseuille-lambda} additionally show the analytical result from ref.~\citenum{waters70a} for comparison.
The agreement with the analytics can be improved to around $1\%$ in all cases if $L$ is used as a fit parameter.
This is justified by the fact that the boundary position in LB is not guaranteed to be exactly at the edge of the cell \cite{ginzbourg94a} and that the extrapolation of \cref{eq:extrapolation} introduces an error for the FV method.
The resulting $L$ differs from the input parameter by $\pm0.6$ cells, or $\pm0.3$ per boundary,
well within the range expected for regular LB.

\subsection{Steady shear flow}
\label{subsec:couette}

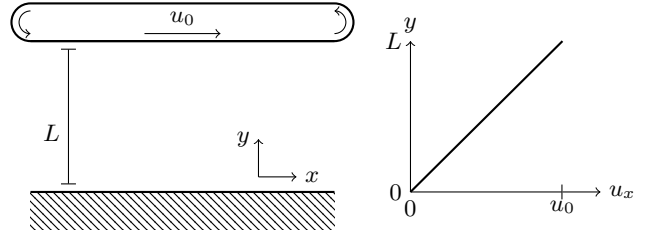
\begin{figure}[tb]
\centering
\tikzsetnextfilename{couette}
\begin{tikzpicture}
	\fill[pattern=north west lines] (0,-0.5) rectangle (4,0);
	\draw[thick] (0,0) -- (4,0);
	\draw[thick] (0,2) -- (4,2);
	\draw[thick] (0,2.5) arc (90:270:0.25) -- (4,2) arc (-90:90:0.25) -- cycle;
	\draw[thin,->] (0,2.4) arc(90:270:0.15);
	\draw[thin,->] (4,2.1) arc(-90:90:0.15);
	\draw[thin,->] (1.5,2.1) -- node[anchor=south] {$u_0$} (2.5,2.1);
	\draw[|-|] (0.5,0.1) -- node[anchor=north east] {$L$} (0.5,1.9);

	\draw[->] (3,0.2) -- (3,0.7) node[anchor=east] {$y$};
	\draw[->] (3,0.2) -- (3.5,0.2) node[anchor=west] {$x$};

	\begin{scope}[xshift=5cm,yshift=1cm]
		\draw[thick,variable=\y, domain=-1:1] plot ({\y+1},{\y});
		\draw[-|] (0,-1) node[anchor=north] {$0$} -- (2,-1) node[anchor=north] {$u_0$};
		\draw[->] (2,-1) -- (2.5,-1) node[anchor=west] {$u_x$};
		\draw[->] (0,-1) node[anchor=east] {$0$} -- (0,1) node[anchor=south] {$y$} node[anchor=east] {$L$};
	\end{scope}
\end{tikzpicture}
\caption{
Geometry of the planar Couette flow system.
A velocity boundary condition of $u_0$ is applied to one side of a fluid in a periodic channel of width $L$,
which leads to a linear profile across the channel for the flow velocity $u_x$ along the channel.
}
\label{fig:couette}
\end{figure}

The planar Couette geometry is similar to that of \cref{subsec:poiseuille},
but replaces the applied force with a velocity boundary condition of $\vec{u}((x,L)^\intercal,t)=u_0\unitvec{e}_x$ on one of the planes,
as illustrated in \cref{fig:couette}.
This relative motion leads to a linear steady-state velocity profile across the channel.
The first normal stress difference,
\begin{equation}
	N_1=\tau_{xx}-\tau_{yy}=2\eta_\text{p}\lambda_\text{p}\frac{u_0^2}{L^2},
\end{equation}
is expected to be constant over the entire channel,
as obtained by solving \cref{eq:oldroydb} with the given velocity profile.
We choose the channel width $L=28\agrid$, the applied velocity $u_0=10^{-3}\agrid/\tgrid$, Newtonian viscosity $\eta_\text{n}=\rho\agrid^2/\tgrid-\eta_\text{p}$, polymer viscosity ratios $\viscosityfraction\in\{0.2,\allowbreak 0.4,\allowbreak 0.6,\allowbreak 0.8\}$, and polymer relaxation times $\lambda_\text{p}/\tgrid\in [1000,\allowbreak 20000]$ for our test simulations.
They are run until sufficiently converged,
which we find to be the case at $t=10\lambda_\text{p}$.
We find that $N_1$ agrees with the prediction to within $0.2\%$ across all parameters.
Appreciable deviations ($\sim 5\%$) are only seen in the cells directly at the boundaries,
where this is expected due to the the stress extrapolation of \cref{eq:extrapolation}.

\subsection{Lid-driven cavity}
\label{subsec:ldc}

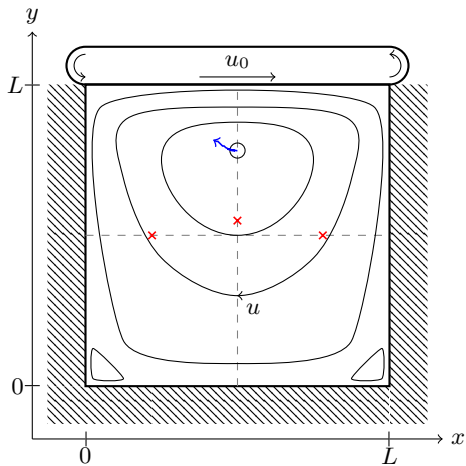
\begin{figure}[tb]
\centering
\tikzsetnextfilename{ldc}
\begin{tikzpicture}
	\begin{scope}
		\clip (4.5,4) rectangle (4,-0.5) rectangle (-0.5,0) rectangle (0,4) ;
		\fill[pattern=north west lines] (-0.5,-0.5) rectangle (4.5,4.5);
	\end{scope}
	\draw[thick] (0,0) rectangle (4,4);
	\draw[thick] (0,4.5) arc (90:270:0.25) -- (4,4) arc (-90:90:0.25) -- cycle;
	\draw[thin,->] (0,4.4) arc(90:270:0.15);
	\draw[thin,->] (4,4.1) arc(-90:90:0.15);
	\draw[thin,->] (1.5,4.1) -- node[anchor=south] {$u_0$} (2.5,4.1);
	
	\draw (-0.7,-0.7) -- (0,-0.7) node[anchor=north] {$0$};
	\draw[|-|] (0,-0.7) -- (4,-0.7) node[anchor=north] {$L$};
	\draw[->] (4,-0.7) -- (4.7,-0.7) node[anchor=west] {$x$} ;
	\draw (-0.7,-0.7) -- (-0.7,0) node[anchor=east] {$0$};
	\draw[|-|] (-0.7,0) -- (-0.7,4) node[anchor=east] {$L$};
	\draw[->] (-0.7,4) -- (-0.7,4.7) node[anchor=south] {$y$} ;
	
	\draw[dashed,black!50] (0,2) -- (4,2);
	\draw[dashed,black!50] (2,0) -- (2,4);
	
	\draw (2,3.124) circle(0.1);
	
	\begin{scope}[xshift=2cm,yshift=2cm]
		\draw plot[smooth cycle, tension=1] coordinates { (-1,1) (0,1.5) (1,1) (0,0) } ;
		\draw plot[smooth cycle, tension=0.6] coordinates { (-1.5,0.75) (-1.45,1.55) (0,1.7) (1.45,1.55) (1.5,0.75) (1,-0.3) (0,-0.8) (-1,-0.3) } ;
		\draw[<-] (0,-0.8) -- node[anchor=north west] {$u$} (0.001,-0.8);
		\draw plot[smooth cycle, tension=0.35] coordinates { (-1.8,1.7) (1.8,1.7) (1.5,-1.4) (0,-1.7) (-1.5,-1.4) };
		\draw plot[smooth cycle, tension=0.35] coordinates { (-1.9,-1.9) (-1.5,-1.9) (-1.9,-1.5) };
		\draw plot[smooth cycle, tension=0.35] coordinates { (1.9,-1.9) (1.5,-1.9) (1.9,-1.5) };
	\end{scope}
	
	\begin{axis}[xmin=0,xmax=1,ymin=0,ymax=1, width=5.58cm,height=5.58cm, ticks=none, hide axis, semithick]
		\addplot[->,blue] table [x index={1}, y index={2}, col sep=space, mark=none] {plots/data/ldc.csv};
		\addplot[draw=none,red,mark=x] coordinates { (0.5, 0.549) (0.219, 0.5) (0.781, 0.5) };
	\end{axis}
\end{tikzpicture}
\caption{
Geometry of the lid-driven cavity system.
A square flow cell of size $L$ has no-slip boundaries on three sides and a constant-velocity boundary condition of $u_0$ along the fourth.
The resulting flow develops a primary vortex near the top middle of the cell.
Along the dashed lines, flow velocity minima and maxima are found at the red crosses.
The blue arrow indicates how the vortex center moves as $\mathrm{Wi}$ is increased from 0 to 1.
}
\label{fig:ldc}
\end{figure}

The lid-driven cavity consists of a square flow cell of edge length $L$, with no-slip boundaries on three sides and a constant velocity boundary $\vec{u}((x,L)^\intercal,t)=u_0\unitvec{e}_x$ on the top side.
This is depicted in \cref{fig:ldc}, which also illustrates the shape of the resulting flow:
a primary vortex develops near the top center of the flow cell and secondary vortices arise in the lower corners.
The exact position of the center of the primary vortex, as well as the position $y$ of the minimum of $u_x((L/2,y)^\intercal,\infty)$ and the positions $x$ of the minimum and maximum of $u_y((x,L/2)^\intercal,\infty)$ vary with the flow parameters and have been extensively studied in literature \cite{yapici09a,sousa16a,dalal16a,pan09a,habla14a}, making them well-suited for comparison in the following.

\begin{figure}[tb]
\centering
\tikzsetnextfilename{ldc-pos}
\input{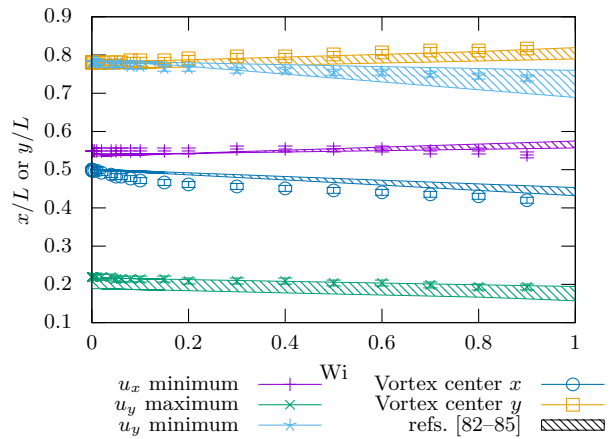}
\caption{
Positions of the primary vortex and flow extrema in the lid-driven cavity.
Colors refer to the different points.
Symbols are our results, while the hatched areas indicates the range covered by the numerical results from refs.~\citenum{sousa16a,dalal16a,pan09a,habla14a}.
}
\label{fig:ldc-pos}
\end{figure}

\begin{figure}[tb]
\centering
\tikzsetnextfilename{ldc-u}
\begin{tikzpicture}[gnuplot]
\tikzset{every node/.append style={scale=0.88}}
\path (0.000,0.000) rectangle (8.000,6.000);
\gpcolor{color=gp lt color border}
\gpsetlinetype{gp lt border}
\gpsetdashtype{gp dt solid}
\gpsetlinewidth{1.00}
\draw[gp path] (1.316,1.674)--(1.496,1.674);
\draw[gp path] (7.516,1.674)--(7.336,1.674);
\node[gp node right] at (1.155,1.674) {-0.2};
\draw[gp path] (1.316,2.688)--(1.496,2.688);
\draw[gp path] (7.516,2.688)--(7.336,2.688);
\node[gp node right] at (1.155,2.688) {-0.1};
\draw[gp path] (1.316,3.702)--(1.496,3.702);
\draw[gp path] (7.516,3.702)--(7.336,3.702);
\node[gp node right] at (1.155,3.702) {0.0};
\draw[gp path] (1.316,4.715)--(1.496,4.715);
\draw[gp path] (7.516,4.715)--(7.336,4.715);
\node[gp node right] at (1.155,4.715) {0.1};
\draw[gp path] (1.316,5.729)--(1.496,5.729);
\draw[gp path] (7.516,5.729)--(7.336,5.729);
\node[gp node right] at (1.155,5.729) {0.2};
\draw[gp path] (1.316,1.674)--(1.316,1.854);
\draw[gp path] (1.316,5.729)--(1.316,5.549);
\node[gp node center] at (1.316,1.404) {$0$};
\draw[gp path] (2.556,1.674)--(2.556,1.854);
\draw[gp path] (2.556,5.729)--(2.556,5.549);
\node[gp node center] at (2.556,1.404) {$0.2$};
\draw[gp path] (3.796,1.674)--(3.796,1.854);
\draw[gp path] (3.796,5.729)--(3.796,5.549);
\node[gp node center] at (3.796,1.404) {$0.4$};
\draw[gp path] (5.036,1.674)--(5.036,1.854);
\draw[gp path] (5.036,5.729)--(5.036,5.549);
\node[gp node center] at (5.036,1.404) {$0.6$};
\draw[gp path] (6.276,1.674)--(6.276,1.854);
\draw[gp path] (6.276,5.729)--(6.276,5.549);
\node[gp node center] at (6.276,1.404) {$0.8$};
\draw[gp path] (7.516,1.674)--(7.516,1.854);
\draw[gp path] (7.516,5.729)--(7.516,5.549);
\node[gp node center] at (7.516,1.404) {$1$};
\draw[gp path] (1.316,5.729)--(1.316,1.674)--(7.516,1.674)--(7.516,5.729)--cycle;
\node[gp node left] at (4.000,3.000) {$\phantom{\ref{sec:theory}}$};
\node[gp node center,rotate=-270] at (0.255,3.701) {$u_x/u_0$ or $u_y/u_0$};
\node[gp node center] at (4.416,0.999) {$\mathrm{Wi}$};
\gpcolor{rgb color={0.580,0.000,0.827}}
\gpsetdashtype{gp dt 1}
\draw[gp path] (1.316,2.586)--(3.796,2.738)--(6.276,2.860)--(7.516,2.916);
\gpcolor{rgb color={0.000,0.620,0.451}}
\draw[gp path] (1.316,4.685)--(3.796,4.594)--(6.276,4.523)--(7.516,4.507);
\gpcolor{rgb color={0.337,0.706,0.914}}
\draw[gp path] (1.316,2.708)--(3.796,2.860)--(6.276,3.022)--(7.516,3.073);
\gpcolor{rgb color={0.580,0.000,0.827}}
\gpsetpointsize{6.00}
\gppoint{gp mark 1}{(1.316,2.492)}
\gppoint{gp mark 1}{(1.317,2.492)}
\gppoint{gp mark 1}{(1.317,2.492)}
\gppoint{gp mark 1}{(1.319,2.492)}
\gppoint{gp mark 1}{(1.322,2.492)}
\gppoint{gp mark 1}{(1.328,2.492)}
\gppoint{gp mark 1}{(1.347,2.492)}
\gppoint{gp mark 1}{(1.378,2.492)}
\gppoint{gp mark 1}{(2.246,2.463)}
\gppoint{gp mark 1}{(1.440,2.491)}
\gppoint{gp mark 1}{(1.564,2.489)}
\gppoint{gp mark 1}{(1.626,2.488)}
\gppoint{gp mark 1}{(1.688,2.486)}
\gppoint{gp mark 1}{(1.812,2.482)}
\gppoint{gp mark 1}{(1.936,2.477)}
\gppoint{gp mark 1}{(2.556,2.452)}
\gppoint{gp mark 1}{(3.176,2.447)}
\gppoint{gp mark 1}{(3.796,2.463)}
\gppoint{gp mark 1}{(4.416,2.492)}
\gppoint{gp mark 1}{(5.036,2.531)}
\gppoint{gp mark 1}{(5.656,2.577)}
\gppoint{gp mark 1}{(6.276,2.628)}
\gppoint{gp mark 1}{(6.896,2.697)}
\gpcolor{rgb color={0.000,0.620,0.451}}
\gppoint{gp mark 2}{(1.316,4.764)}
\gppoint{gp mark 2}{(1.317,4.764)}
\gppoint{gp mark 2}{(1.317,4.764)}
\gppoint{gp mark 2}{(1.319,4.764)}
\gppoint{gp mark 2}{(1.322,4.764)}
\gppoint{gp mark 2}{(1.328,4.765)}
\gppoint{gp mark 2}{(1.347,4.766)}
\gppoint{gp mark 2}{(1.378,4.768)}
\gppoint{gp mark 2}{(2.246,4.827)}
\gppoint{gp mark 2}{(1.440,4.772)}
\gppoint{gp mark 2}{(1.564,4.780)}
\gppoint{gp mark 2}{(1.626,4.784)}
\gppoint{gp mark 2}{(1.688,4.789)}
\gppoint{gp mark 2}{(1.812,4.798)}
\gppoint{gp mark 2}{(1.936,4.807)}
\gppoint{gp mark 2}{(2.556,4.842)}
\gppoint{gp mark 2}{(3.176,4.855)}
\gppoint{gp mark 2}{(3.796,4.854)}
\gppoint{gp mark 2}{(4.416,4.844)}
\gppoint{gp mark 2}{(5.036,4.830)}
\gppoint{gp mark 2}{(5.656,4.814)}
\gppoint{gp mark 2}{(6.276,4.795)}
\gppoint{gp mark 2}{(6.896,4.765)}
\gpcolor{rgb color={0.337,0.706,0.914}}
\gppoint{gp mark 3}{(1.316,2.639)}
\gppoint{gp mark 3}{(1.317,2.639)}
\gppoint{gp mark 3}{(1.317,2.639)}
\gppoint{gp mark 3}{(1.319,2.639)}
\gppoint{gp mark 3}{(1.322,2.639)}
\gppoint{gp mark 3}{(1.328,2.640)}
\gppoint{gp mark 3}{(1.347,2.640)}
\gppoint{gp mark 3}{(1.378,2.642)}
\gppoint{gp mark 3}{(2.246,2.655)}
\gppoint{gp mark 3}{(1.440,2.645)}
\gppoint{gp mark 3}{(1.564,2.650)}
\gppoint{gp mark 3}{(1.626,2.652)}
\gppoint{gp mark 3}{(1.688,2.654)}
\gppoint{gp mark 3}{(1.812,2.656)}
\gppoint{gp mark 3}{(1.936,2.657)}
\gppoint{gp mark 3}{(2.556,2.653)}
\gppoint{gp mark 3}{(3.176,2.657)}
\gppoint{gp mark 3}{(3.796,2.678)}
\gppoint{gp mark 3}{(4.416,2.709)}
\gppoint{gp mark 3}{(5.036,2.748)}
\gppoint{gp mark 3}{(5.656,2.790)}
\gppoint{gp mark 3}{(6.276,2.833)}
\gppoint{gp mark 3}{(6.896,2.887)}
\gpcolor{color=gp lt color border}
\node[gp node right] at (3.270,0.855) {$u_x$ minimum};
\gpcolor{rgb color={0.580,0.000,0.827}}
\gpsetdashtype{gp dt solid}
\draw[gp path] (3.431,0.855)--(4.255,0.855);
\gppoint{gp mark 1}{(3.843,0.855)}
\gpcolor{color=gp lt color border}
\node[gp node right] at (3.270,0.585) {$u_y$ maximum};
\gpcolor{rgb color={0.000,0.620,0.451}}
\draw[gp path] (3.431,0.585)--(4.255,0.585);
\gppoint{gp mark 2}{(3.843,0.585)}
\gpcolor{color=gp lt color border}
\node[gp node right] at (3.270,0.315) {$u_y$ minimum};
\gpcolor{rgb color={0.337,0.706,0.914}}
\draw[gp path] (3.431,0.315)--(4.255,0.315);
\gppoint{gp mark 3}{(3.843,0.315)}
\gpcolor{rgb color={0.580,0.000,0.827}}
\gppoint{gp mark 4}{(1.316,2.001)}
\gpcolor{rgb color={0.000,0.620,0.451}}
\gppoint{gp mark 4}{(1.316,5.195)}
\gpcolor{rgb color={0.337,0.706,0.914}}
\gppoint{gp mark 4}{(1.316,2.208)}
\gpcolor{color=gp lt color border}
\node[gp node right] at (6.670,0.855) {Newtonian};
\gppoint{gp mark 4}{(7.243,0.855)}
\node[gp node right] at (6.670,0.585) {ref.~\cite{habla14a}};
\gpsetdashtype{gp dt 1}
\draw[gp path] (6.831,0.585)--(7.655,0.585);
\gpsetdashtype{gp dt solid}
\draw[gp path] (1.316,5.729)--(1.316,1.674)--(7.516,1.674)--(7.516,5.729)--cycle;
\gpdefrectangularnode{gp plot 1}{\pgfpoint{1.316cm}{1.674cm}}{\pgfpoint{7.516cm}{5.729cm}}
\end{tikzpicture}
\caption{
Velocity of the flow extrema in the lid-driven cavity.
Colors refer to the different points.
Symbols are our results, while the line refers to numerical results from ref.~\citenum{habla14a}.
The square symbols indicate Newtonian simulations ($\viscosityfraction\rightarrow 0$) and match results from ref.~\citenum{sousa16a}, while the others are viscoelastic ($\viscosityfraction=0.5$).
}
\label{fig:ldc-u}
\end{figure}
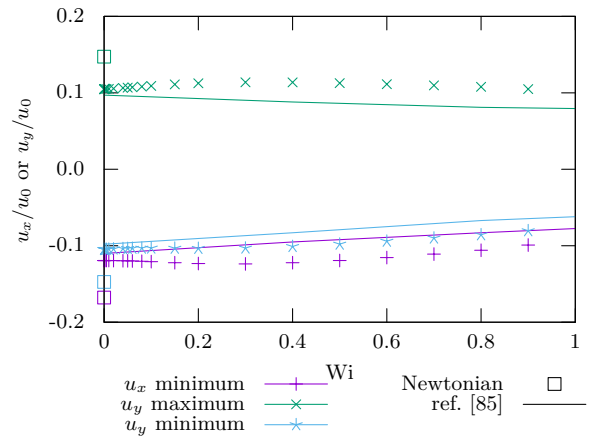

We choose width and height $L=194$ for the flow cell, Newtonian viscosity $\eta_\text{n}=\rho\agrid^2/\tgrid-\eta_\text{p}$, applied velocity $u_0=10^{-4}\agrid/\tgrid$, polymer viscosity ratio $\viscosityfraction=0.5$, and polymer relaxation times $\lambda_\text{p}$ such that Weissenberg numbers $\mathrm{Wi}\in [0,1]$ are obtained.
For $\mathrm{Wi}=0$, $\viscosityfraction\rightarrow 0$ is also used.
The Weissenberg and Deborah numbers coincide as \cite{sousa16a}
\begin{equation}
	\mathrm{Wi}=\mathrm{De}=\frac{\lambda_\text{p}u_0}{L}
\end{equation}
for the system under consideration.
The Reynolds number is given by
\begin{equation}
	\mathrm{Re}=\frac{\rho u_0L}{\totalviscosity} = 0.02,
\end{equation}
again placing us in the low-Reynolds regime.

For numerical reasons, the velocity boundary condition is not applied as given above. Instead, a regularization is used to remove the infinite flow divergence in the top corners.
A common choice is
\begin{equation}
	\vec{u}((x,L)^\intercal,t)=16u_0\left(\frac{x}{L}\right)^2\left(1-\frac{x}{L}\right)^2\unitvec{e}_x.
\end{equation}
This regularization leaves the qualitative flow features untouched, but thwarts quantitative comparison with the unregularized simulations of ref.~\citenum{yapici09a}.
The same regularization is employed by refs.~\citenum{sousa16a,dalal16a,pan09a,habla14a} and shall be used in the comparison below.

\Cref{fig:ldc-pos} shows the positions of the primary vortex and the flow velocity extrema in our simulations. Error bars correspond to the size of a cell plus the potential deviation of the true boundary position from the prescribed boundary position.
One can see that the general trend from refs.~\citenum{sousa16a,dalal16a,pan09a,habla14a} is recovered semi-quantitatively, with the exception of the nonlinear deviation of the $x$-component of the vortex center.
Results vary significantly between these references, so that a quantitative comparison is not drawn.
However, in view of this, the result in \cref{fig:ldc-pos} gives confidence in our method's accuracy.
The speed with which our results were obtained,
as well as the ability to refine these significantly,
provide opportunities for future benchmarking.

The flow velocity at the points of interest is shown in \cref{fig:ldc-u}.
Values differ between refs.~\citenum{sousa16a,dalal16a,pan09a,habla14a} by factors of up to $2$,
so we only plot the comparison to ref.~\citenum{habla14a}.
This reference has matching flow velocities at $\mathrm{Wi}\rightarrow 0$ and exhibits the same trend of decreasing velocity magnitudes as our results.
The vortex is observed to move toward the top left as $\mathrm{Wi}$ is increased.
The minimum of $u_x$ moves down slightly,
while both the minimum and the maximum of $u_y$ move toward the left.
The deviations from the results in literature are expected
as the system is very sensitive to resolution,
especially at larger $\mathrm{Wi}$.
Our resolution was chosen such that the results had sufficiently converged.

We also performed one simulation at $\viscosityfraction\rightarrow 0$, the Newtonian case, and observe that this yields a different velocity than $\mathrm{Wi}\rightarrow 0$ at constant $\viscosityfraction=0.5$.
The velocity obtained in the former way agrees with that reported by ref.~\citenum{sousa16a} to within $1\%$.
The latter way corresponds to the case of instantaneous polymer relaxation, but not vanishing viscoelasticity.

\subsection{Four-roll mill}
\label{subsec:4rm}

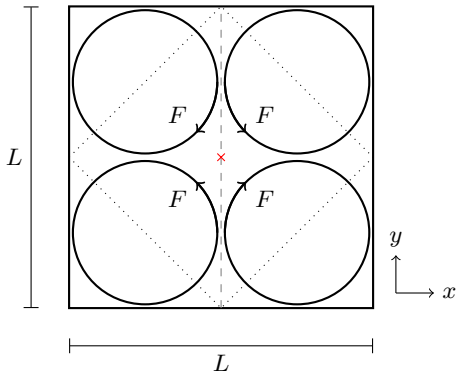
\begin{figure}[tb]
\centering
\tikzsetnextfilename{4rm}
\begin{tikzpicture}
	\pgfmathsetmacro\r{0.95}

	\draw[thick] (0,0) rectangle (4,4);
	\draw[dotted] (0,2) -- (2,4) -- (4,2) -- (2,0) -- (0,2);

	\draw[thick] (1,1) circle (\r);
	\draw[thick] (3,1) circle (\r);
	\draw[thick] (1,3) circle (\r);
	\draw[thick] (3,3) circle (\r);
	\draw[thick,->] (1+\r,3) arc (0:-45:\r) node[anchor=south east] {$F$};
	\draw[thick,->] (3-\r,3) arc (180:225:\r) node[anchor=south west] {$F$};
	\draw[thick,->] (1+\r,1) arc (0:45:\r) node[anchor=north east] {$F$};
	\draw[thick,->] (3-\r,1) arc (180:135:\r) node[anchor=north west] {$F$};

	\draw[dashed,black!50] (2,0) -- (2,4);
	
	\draw[|-|] (-0.5,0) -- node[anchor=east] {$L$} (-0.5,4);
	\draw[|-|] (0,-0.5) -- node[anchor=north] {$L$} (4,-0.5);
	
	\draw[->] (4.3,0.2) -- (4.3,0.7) node[anchor=south] {$y$};
	\draw[->] (4.3,0.2) -- (4.8,0.2) node[anchor=west] {$x$};
	
	\node[red] at (2,2) {\pgfuseplotmark{x}};
\end{tikzpicture}
\caption{
Geometry of the four-roll mill.
Four counter-rotating forces $F$ are applied to a periodic square flow cell of size $L$.
This leads to a pure extensional flow at the center of the cell.
Velocity and stress will be measured along the dashed line.
The dotted square indicates the actual simulation domain used,
which still obeys the periodic boundary conditions.
}
\label{fig:4rm}
\end{figure}

The four-roll mill consists of a square cell with length $L$ and periodic boundary conditions.
A force field of
\begin{equation}
	\vec{F}(\vec{r},t)=\frac{8\pi^2\eta_\text{n}u_0}{L^2} \begin{pmatrix}
		\sin\left(\frac{2\pi}{L}x\right) \cos\left(\frac{2\pi}{L}y\right) \\[1em]
		\cos\left(\frac{2\pi}{L}x\right) \sin\left(\frac{2\pi}{L}y\right)
	\end{pmatrix}
\end{equation}
is applied to it, resulting in four counter-rotating rolls as illustrated in \cref{fig:4rm}.
Reference~\citenum{thomases07a} provides an analytical prediction for the steady-state stress in the vicinity of the central point,
where the flow is purely extensional, i.e., $\vec{u}((L/2,L/2)^\intercal,t)=\alpha(\unitvec{e}_x-\unitvec{e}_y)$.

We choose cell size $L=214\sqrt{2}\agrid$,
Newtonian viscosity $\eta_\text{n}=1.5\rho\agrid^2/\tgrid$, polymer viscosity ratio $\viscosityfraction=\frac{1}{3}$,
maximum velocity $u_0=10^{-3}\agrid/\tgrid$,
and polymer relaxation times $\lambda_\text{p}/\tgrid\in\left[1000,\allowbreak  24000\right]$.
The simulation is run until sufficiently converged,
which we find to be the case at $t=20\lambda_\text{p}$.
The Weissenberg number is given by \cite{thomases07a,pimenta17a}
\begin{equation}
	\mathrm{Wi}=\frac{4\pi\lambda_\text{p}u_0}{L},
\end{equation}
and the Reynolds number is low at
\begin{equation}
	\mathrm{Re}=\frac{\rho u_0L}{\totalviscosity}=0.1.
\end{equation}

We found that our simulations lead to a decoupling of the stress at the center point from the rest of the domain due to the upwind scheme from \cref{subsubsec:stability}.
To avoid this, we rotated the lattice by $45^\circ$ relative to the system as indicated in \cref{fig:4rm},
while ensuring that the periodic continuation of the system remains intact.
We would like to stress that this is a rather unusual situation, which only appears here
due to the high level of symmetry and the divergence at the central point.
Such behavior will not commonly appear in soft matter systems,
but when it does, it is easily identified in the stress profiles.
This gives users a means to eliminate potentially problematic simulation runs.

\begin{figure}[tb]
\centering
\tikzsetnextfilename{4rm-stress}
\input{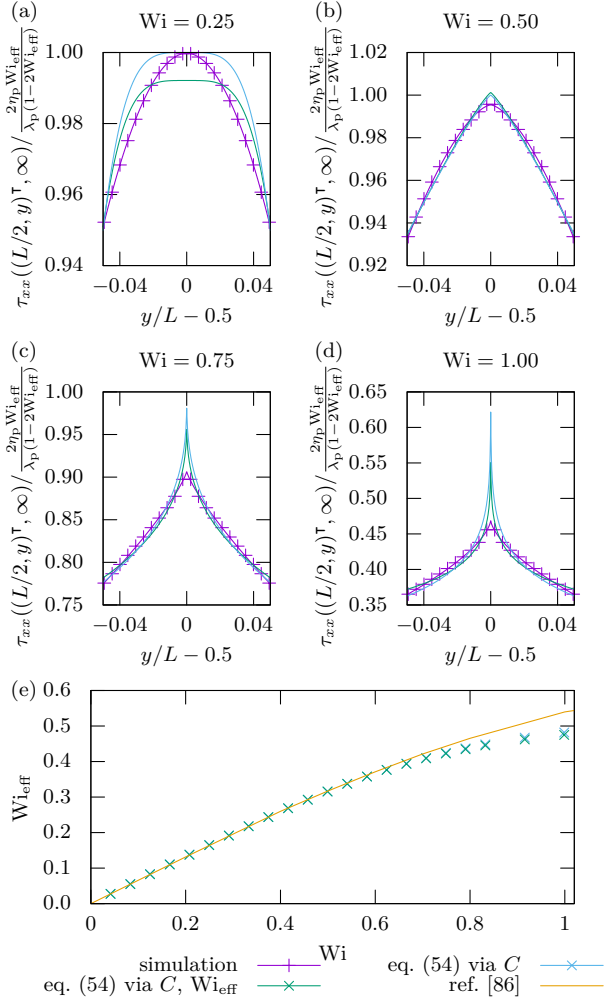}
\caption{(a-d) Stress $\tau_{xx}((L/2,y)^\intercal,\infty)$ near the center of the four-roll mill for different polymer relaxation times $\lambda_\text{p}$.
Symbols are our results, with their connecting line coming from fitting \cref{eq:4rmstress} with an added offset.
The other lines are fits with \cref{eq:4rmstress} via one or two parameters.
(e) $\mathrm{Wi}_\text{eff}$ plotted over the Weissenberg number $\mathrm{Wi}$.
Symbols are our results, while the line comes from ref.~\citenum{thomases07a}.
}
\label{fig:4rm-stress}
\end{figure}

\Cref{fig:4rm-stress}a-d shows the stress component $\tau_{xx}$ along a vertical line through the center of the cell, which is marked with a red cross in \cref{fig:4rm}.
A comparison with ref.~\citenum{thomases07a} is drawn by fitting with its local solution,
\begin{equation}
	\tau_{xx}((L/2,y)^\intercal,\infty)=\frac{2\eta_\text{p}\mathrm{Wi}_\text{eff}}{\lambda_\text{p}(1-2\mathrm{Wi}_\text{eff})}+C\left|\frac{y}{L}-\frac{1}{2}\right|^{\frac{1-2\mathrm{Wi}_\text{eff}}{\mathrm{Wi}_\text{eff}}}.
\label{eq:4rmstress}
\end{equation}
We fit via $C$ while keeping $\mathrm{Wi}_\text{eff}=\lambda_\text{p}\alpha$ constant, as well as via both $C$ and $\mathrm{Wi}_\text{eff}$.
We find that $\mathrm{Wi}_\text{eff}$ only differs by less than $1\%$ between the two fits,
yet the latter fit is significantly better.
This is because fitting an exponent is very sensitive to small deviations.
For $\mathrm{Wi}_\text{eff}<1/4$, the structure of the stress profile is not captured well by the fit.
This is due to the lack of a singularity, as \cref{eq:4rmstress} was constructed with a singularity in mind \cite{thomases07a}.
Beyond this value, three regimes of solutions are recovered:
continuous and differentiable at the center ($\mathrm{Wi}_\text{eff}< 1/3$),
continuous but not differentiable at the center ($1/3\le\mathrm{Wi}_\text{eff}< 1/2$),
and diverging at the center ($\mathrm{Wi}_\text{eff}> 1/2$).
We reproduce the expected regimes,
albeit with the caveat that divergences in our scheme are not present,
due to the smoothing of solutions that its discretization imposes. 
\Cref{fig:4rm-stress}e plots $\mathrm{Wi}_\text{eff}$ that we obtained from the fits via $\mathrm{Wi}$.
Comparison with the corresponding plot from ref.~\citenum{thomases07a} is excellent up to $\mathrm{Wi}_\text{eff}\approx 0.4$ ($\mathrm{Wi}\approx 0.75$),
as expected due to implicit smoothing of the divergences.

\subsection{Settling sphere}
\label{subsec:sphere}

\begin{figure}[tb]
\centering
\tikzsetnextfilename{sphere}
\begin{tikzpicture}
	\draw[thick] (0,0) rectangle (4,4);
	\fill[pattern=crosshatch dots] (2,2.2) circle (0.8);
	\draw[thick] (2,2.2) circle (0.8);
	
	\draw[->] (2,1.4) -- node[anchor=west] {$F$} (2,0.7);
	
	\draw[->] (2,3) -- (2,3.7);
	\draw[->] (1.826,3.4) arc (150:390:0.2 and 0.15) node[anchor=west] {$M$};
	
	\draw[|-|] (0.5,0.1) -- (0.5,0.5) node[anchor=west] {$L$} -- (0.5,3.9);
	\draw[->] (3.5,0.1) -- (3.5,0.5) node[anchor=east] {$-F$} -- (3.5,0.8);
	\draw[|-|] (3.0,1.4) -- (3.0,2.2) node[anchor=west] {$2R$} -- (3.0,3.0);
	
	\draw[->] (4.3,0.2) -- (4.3,0.7) node[anchor=south] {$z$};
	\draw[->] (4.3,0.2) -- (4.8,0.2) node[anchor=west] {$x$};
\end{tikzpicture}
\caption{
Geometry of the sedimenting sphere system.
A sphere of radius $R$ sediments under velocity $v$ due to an applied force $F$ in a periodic cubic box of length $L$.
A torque $M$ is applied to the sphere to rotate it with velocity $\omega$.
}
\label{fig:sphere}
\end{figure}
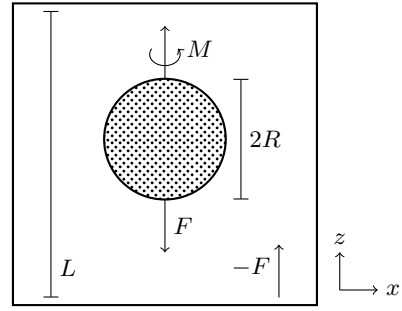

So far, all systems investigated were two-dimensional and had constant boundary conditions.
To demonstrate our algorithm's capabilities beyond this, we simulate the sedimentation of a rotating sphere.
A sphere of radius $R$ is placed in a cubic box of size $L^3$ with periodic boundary conditions.
A constant force $\vec{F}=F_z\unitvec{e}_z$ is applied to the sphere and the counterforce $-\vec{F}$ is distributed evenly among all fluid cells so that the net momentum of the system remains zero.
Furthermore, a constant torque $\vec{M}=M_z\unitvec{e}_z$ is applied to the sphere to rotate it around the $z$-axis; a counter-torque on the fluid is not needed \cite{fischer15a}.
The geometry is illustrated in \cref{fig:sphere}.

\begin{figure}[tb]
\centering
\tikzsetnextfilename{sphere-u}
\input{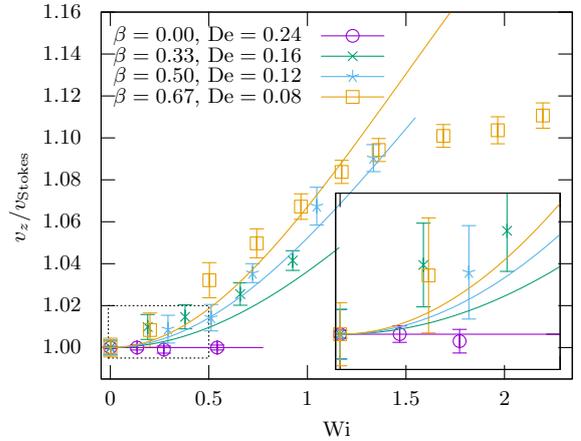}
\caption{
Sedimentation velocity of a rotating sphere in an Oldroyd-B fluid.
Symbols with error bars are our results, while the lines are theoretical predictions from ref.~\citenum{housiadas19a} for the same parameters (ending at $\mathrm{Wi}_\text{max}$ as per \cref{eq:wimax}).
The inset displays an enlargement of the lower left region of the large graph indicated by the dotted box.
}
\label{fig:sphere-u}
\end{figure}

We choose our parameters as $R=8\agrid$, $L/R\in [7.5,\allowbreak 30]$, $F_z=0.008\rho\agrid^4/\tgrid^2$, $\eta_\text{n}=\frac{1}{6}\rho\agrid^2/\tgrid$, $\eta_\text{p}/\eta_\text{n}\in\{0,\allowbreak \frac{1}{2},\allowbreak 1,\allowbreak 2\}$ and $\lambda_\text{p}=6000\tgrid$.
The simulation is run until the velocity $\vec{v}$ of the sphere has converged, for which $t=10\lambda_\text{p}$ tends to suffice.
We can assume $M_z=0$ since it does not change the order of magnitude of the sedimentation velocity $\vec{v}$ \cite{housiadas19a} and employ Stokes' law,
\begin{align}
	v_\text{Stokes}
	&= \frac{F}{6\pi\totalviscosity R},
\end{align}
in order to estimate the Reynolds number for our parameter range as
\begin{align}
	\mathrm{Re}
	&=\frac{2\rho v_\text{Stokes}R}{\totalviscosity}
	=\frac{\rho F}{3\pi\totalviscosity^2}
	\in [0.003,0.03],
\end{align}
which lies well in the low-Reynolds regime.
The Weissenberg and Deborah numbers of the system are given by \cite{housiadas19a}
\begin{align}
	\mathrm{Wi} &= \lambda_\text{p} \omega_z \displaybreak[0]\\
	\mathrm{De} &= \frac{\lambda_\text{p} v_0}{R},
\end{align}
where $\vec{\omega}=\omega_z\unitvec{e}_z$ is the measured angular velocity of the sphere.
$v_0$ is the sedimentation velocity measured for $\omega_z=0$, with all other parameters kept equal.
$\omega_z$ can be varied by changing the applied torque $M_z$.
$M_z$ is chosen such that we cover a range of Weissenberg numbers while staying below a certain value of the tangential velocity $v_\text{t}=\omega_zR$ in order to not jeopardize the LB's stability.
To achieve this, we define a maximum surface Reynolds number
\begin{align}
	\mathrm{Re}_\text{t,max}
	&=\frac{2\rho v_\text{t,max}R}{\totalviscosity}
	\equiv 0.1,
\end{align}
which can be used to obtain a maximum allowed Weissenberg number as
\begin{align}
	\mathrm{Wi}_\text{max}
	&=\lambda_\text{p}\omega_\text{max}
	=\frac{\lambda_\text{p} \mathrm{Re}_\text{t,max}\totalviscosity}{2\rho R^2}.
	\label{eq:wimax}
\end{align}

The parameters provided above correspond to four sets of simulations with different polymer viscosity fractions $\viscosityfraction$.
Within each set, the variation of $\omega_z$ or $M_z$ corresponds to a change in $\mathrm{Wi}$, which makes the horizontal axis of \cref{fig:sphere-u}.
To obtain the value on the vertical axis, first an exponential decay is fitted to $\vec{v}(t)$ to extrapolate to $t\rightarrow\infty$, and then simulations at different $L$ are used to extrapolate it to $L\rightarrow\infty$.
The fit error of these two processes is used to obtain the plot error bars.
In \cref{fig:sphere-u}, we also compare to an analytical solution by \citet{housiadas19a}, who expanded $v/v_\text{Stokes}$ in terms of $\mathrm{De}$ for arbitrary $\viscosityfraction$ and $\chi=\mathrm{Wi}/\mathrm{De}$.
Agreement is mostly within error bars up to $\mathrm{Wi}\approx 1$.
Deviations beyond that are comparable to those found by ref.~\citenum{housiadas19a}'s own comparison to numerical results from ref.~\citenum{castillo19a} for similar parameters.
This shows that our method reproduces the analytical solution in its range of validity,
while behaving similar to other methods beyond that realm.

\IfFileExists{snowman-squirmer.tex}{
	\input{snowman-squirmer.tex}
}{}

\section{Summary and outlook}
\label{sec:summary}

We have introduced a method to simulate Oldroyd-B fluids with lattice Boltzmann.
It uses moving boundaries to allow for the simulation of suspended colloids.
We validated our method against several rheological benchmark problems
and determined it to correspond well with literature for Weissenberg and Deborah numbers and viscosity fractions between zero and one,
a regime relevant for many colloidal systems.
We also validated our method for
\IfFileExists{snowman-squirmer.tex}{
	specific colloidal problems,
	spheres sedimenting under an applied torque and squirmers swimming,
	where analytical and numerical predictions are recovered in their regime of validity.
}{
	a specific colloidal problem,
	a sphere sedimenting under an applied torque,
	where analytical predictions are recovered in their regime of validity.
}
Computational effort scales linearly with the number of fluid cells,
while the computational cost of adding particles is negligible compared to that of simulating the fluid.
Published data on the benchmarks we considered for this work covered only a small parameter space, i.e. the few most relevant points,
therefore we will make our full data set available to serve as a reliable reference for future investigations.
The simulation code will also be provided to enable others to study similar systems at parameters and resolutions of their choosing.
Finally, thanks to the use of automatic code generation, our model and implementation are easily extensible to other viscoelastic models.
Incorporating thermal fluctuations \cite{huetter20a-pre} is also conceivable.

Our viscoelastic, moving-boundary LB facilitates future study of dense colloidal suspensions in viscoelastic fluids.
This might include the collective sedimentation of colloids \cite{murch20a},
which goes beyond the single-body effects discussed in \cref{subsec:sphere}.
The field of self-propelled colloids is of particular interest to us.
Previous reports of viscoelastic enhancement of rotational diffusivity \cite{gomezsolano16a}, for example, have spurred interest in the community.
Simulation studies \cite{qi20a} however could not discern whether this was an effect of viscoelasticity or merely of an inhomogeneous polymer concentration.
Our method does away with the explicit consideration of polymers and might settle such questions.
Besides effective propulsion models \cite{binagia20a,qi20a}, fully-resolved propulsion models \cite{kuron16a} might also be used,
which would permit investigating complex phenomena arising from the interplay of hydrodynamics, viscoelasticity, electrostatics and phoretic interactions, such as those experimentally studied in ref.~\citenum{saad19a}.
Our new and extensively validated method provides a first stepping stone toward such future physical modeling.

\begin{acknowledgement}
  We are grateful to Martin Bauer for help with \textsf{pystencils} and thank Ashreya Jayaram, Alexander Morozov, Becca Thomases, and Rudolf Weeber for useful discussions
  and Fabian H\"ausl for helpful comments on the manuscript.
  We acknowledge the Deutsche Forschungsgemeinschaft (DFG) for funding
  through the SPP 1726 ``Microswimmers: from single particle motion to collective behavior'' (HO1108/24-2)
  and through the EXC 2075 ``SimTech'' (390740016).
  JdG further acknowledges funding by an NWO START-UP grant (740.018.013).
\end{acknowledgement}

\section*{Author contributions}
Conceptualization: MK, JdG;
Calculations and analysis: MK, CS;
Writing: MK, CS, JdG;
Supervision: MK, JdG, CH;
Funding acquisition: JdG, CH;
Resources: CH.

\section*{Research data}
The numerical code and analysis scripts used to obtain the data presented in this publication
are available at \url{https://doi.org/10.24416/UU01-2AFZSW}, along with a representative subset of the data.

\bibliographystyle{jabbrv_epj}
\bibliography{bibtex/icp,paper}

\end{document}